\documentclass[aps,prl,reprint]{revtex4-1}
\usepackage{graphicx}
\usepackage{color}
\usepackage{amsmath}
\usepackage{amssymb}
\usepackage{verbatim,bm}
\def\be{\begin{equation}}
\def\ee{\end{equation}}
\def\ber{\begin{eqnarray}}
\def\eer{\end{eqnarray}}
\def\bern{\begin{eqnarray*}}
\def\eern{\end{eqnarray*}}

\def\rv{{\bf r}}

\def\Gv{{\bf G}}

\def\qv{{\bf q}}

\def\0v{{\bf 0}}

\def\pa{\partial}

\begin{document}
\title{Optics of semiconductors from  meta-GGA-based\\ time-dependent density-functional theory}
\author {V. U. Nazarov}
\affiliation{Research Center for Applied Sciences, Academia Sinica, Taipei 11529, Taiwan}

\author{G. Vignale}
\affiliation{Department of Physics, University of Missouri-Columbia, Columbia, Missouri 65211, USA}
\date{\today}
\begin{abstract}
We calculate the optical spectra of silicon and germanium in the adiabatic time-dependent density functional formalism, making use of  kinetic energy  density-dependent (meta-GGA) exchange-correlation functionals. 
We find excellent agreement between theory and experiment.  The success of the theory on this notoriously difficult problem is traced to the fact that the  exchange-correlation kernel of  meta-GGA supports a singularity of the form $\alpha/q^2$ (where $q$ is the wave-vector and $\alpha$ is a constant), whereas previously employed approximations (e.g. local density  and generalized gradient approximations) do not.   Thus, the use of the adiabatic meta-GGA  opens a new path for handling  the extreme non-locality of the time-dependent exchange-correlation potential in  solid-state systems.
\end{abstract}
\maketitle

The first-principle calculation of the optical properties of semiconductors is a classic and practically important problem in electronic structure theory.   The difficulty stems largely from the critical role played by electron-electron interactions, particularly the so-called excitonic effects, i.e. the interaction of an electron in the conduction band with the hole left behind in the valence band.  Early calculations \cite{Hanke-79,*Hanke-80} based on diagrammatic many-body theory achieved good agreement with the experiment at the price of much computational effort. 
In recent years, the problem has been tackled by several authors \cite{Hybertsen-85,*Albrecht-98,*Ku-02}, 
who made use of state-of-the-art methods such as the GW approximation for the electron self-energy and the Bethe-Salpeter equation for the electron-hole interaction.   These methods are computationally demanding and not so easily adaptable to an emerging new generation of electronic materials, e.g. organic semiconductors and long polymer chains.

A promissing alternative to the traditional many-body approach is provided by the time-dependent density functional theory (TDDFT) \cite{Zangwill-80,*Runge-84,*Gross-85}.  This approach directly targets the density-density (or, in some versions, the current-current \cite{Vignale-96}) response function of a fictitious non-interacting system, the so-called Kohn-Sham (KS) system, which is so designed as to produce (at least in principle)  the same density/current response as the physical interacting system.  The elimination of  interactions greatly reduces the computational effort, but the complexity of the many-body problem eventually resurfaces, since the quality of the results is crucially determined by the quality of the (approximate) exchange-correlation (xc) potential $v_{xc}(\rv,t)$ in which the fictitious non-interacting electrons move.

Successes and failures of the TDDFT approach to the calculation of optical spectra of semiconductor are well-documented.  The first difficulty, which has been known since the early 1980s, is that the basic local-density approximation (LDA) and its semilocal extensions severely underestimate the 
band gap.  The  problem with the {\em KS} band gap can be corrected by the use of orbital-dependent functionals
\cite{Slater-51,*Sharp-53,*Talman-76,Gorling-93,*Gorling-97,Kummel-08}
\footnote{The `true' band gap would require the non-adiabatic TDDFT \cite{Gruning-06},
which lies beyond the scope of this work.} or the TDDFT approach can be implemented on top of a band-structure obtained by a many-body calculation
\cite{Reining-02,Sottile-03,Botti-04}.
However, even if  the band-gap had been corrected, the calculation of the optical properties  is not easy.
The standard approach based on the adiabatic local density approximation (ALDA) \cite{Zangwill-80,Gross-85}, for example, dramatically underestimates the low energy peak -- commonly referred to as the ``excitonic peak" -- in the optical spectrum.   Improvements on the ALDA such as the adiabatic extension of the GGA  do not fare much better. 
The exact exchange approach~\cite{Gorling-93,*Gorling-97} has been found to  
catch the excitonic effect in silicon \cite{Kim-02-2}, but this was achieved at the cost of artificially restricting the set of states included in the calculation  to avoid the ``collapse" of the spectra. So far, the most consistent ab-initio scheme leading  to results in good agreement with experiment has been the recasting of Bethe-Salpeter equation as an equation for a two-point function within the framework of TDDFT \cite{Sottile-03};
but even this approach remains computationally very demanding.

In recent years, a new class of approximate functionals has emerged in ground-state DFT.  These are known as meta-GGA (MGGA) functionals and their defining characteristic is to depend not only on the density and its gradient, but also on the non-interacting kinetic energy density $\tau(\rv)$ \cite{Voorhis-98,Voorhis-08,Tao-03,Becke-06,Tran-09}.   At first sight, the dependence on $\tau(\rv)$ seems to contradict the general statement that the xc potential is a functional of the density.   But, it must be kept in mind that $\tau(\rv)$ is determined by  Kohn-Sham orbitals which, in turn, are nonlocal functionals of the density.  Thus the MGGA functionals are still functionals of the density, but intrinsically nonlocal ones.  Their power stems entirely from this fact.

In this Letter we show that an adiabatic approximation based on meta-GGA functionals leads to very significant improvement in the calculation of optical properties.   The fact that meta-GGA functionals can lead to improvements in the calculation of the KS band gap has been known for some time \cite{Tran-09}.  
What we  add here to that knowledge is the realization that these functionals can also produce  accurate optical spectra.  And since the use of the adiabatic approximation automatically excludes retardation effects, we conclude that the primary reason for the success of the meta-GGA functionals is the improved treatment of the  long-rangedness in the xc potential.  This long-rangedness  (often referred to as ``ultranonlocality") has long been known to be a problem in TDDFT, especially so in the applications to extended systems. 
While its strength could be inferred from  fits to experimental spectra \cite{Reining-02}, none of the approximations developed so far could deal with it satisfactorily.    We  believe that the use of  meta-GGA functionals is a breakthrough in the handling of ultranonlocality and paves the way to  efficient first-principle calculations of the optical properties of semiconductors and more complex materials.  

{\it Formulation --} The crucial quantity targeted in  TDDFT is the density-density response function $\chi(\rv,\rv',\omega)$, which is related to the non-interacting
KS response function $\chi_s(\rv,\rv',\omega)$ by the equation~\cite{Gross-85}
\begin{equation}
 \chi^{-1}(\rv,\rv',\omega)=\chi_{s}^{-1}(\rv,\rv',\omega)-f_{xc}(\rv,\rv',\omega)-\frac{e^2}{|\rv-\rv'|},
\label{chi}
\end{equation}
where 
$f_{xc}(\rv,\rv',\omega)= \delta v_{xc}(\rv,\omega)/\delta n(\rv',\omega)$
is the xc kernel, defined as the functional derivative of the dynamic xc potential $v_{xc}(\rv,\omega)$ with respect to the dynamic particle-density.
In order to calculate $f_{xc}$ we start from the expression for the xc energy within MGGA as
\begin{equation}
 E_{xc}=\int \epsilon_{xc}[n(\rv),\nabla n(\rv),\tau(\rv)] d\rv,
\label{Exc}
\end{equation}
where the xc energy density $\epsilon_{xc}$ is a local function of its three arguments,
$n(\rv)$ is the particle density, 
\begin{eqnarray}
\tau(\rv)&=&\frac{1}{2} \sum\limits_\alpha f_\alpha |\nabla \psi_\alpha(\rv)|^2\cr\cr
      &=&\sum\limits_\alpha f_\alpha \epsilon_\alpha |\psi_\alpha(\rv)|^2 -v_s(\rv) n(\rv) +\frac{1}{4} \nabla^2 n(\rv)
\label{t}
\end{eqnarray}
is the non-interacting kinetic energy density,   $\psi_\alpha$, $\epsilon_\alpha$, and $f_\alpha$ are the KS orbitals, their eigenenergies, and the occupation numbers,
respectively, and $v_s(\rv)$ is the static KS potential.    The second equality in Eq.~(\ref{t}) follows from the KS equation.
With the use of the definitions of the xc potential and the xc kernel as the first and the second functional
derivatives of $E_{xc}$ with respect to density, we derive from Eq.~(\ref{Exc})
\begin{equation}
v_{xc}(\rv)=\frac{\pa \epsilon_{xc}}{\pa n}(\rv)
\! - \!  \nabla \frac{\pa \epsilon_{xc}}{\pa \nabla n} (\rv)
 \! + \! \int \frac{\pa \epsilon_{xc}}{\pa \tau}(\rv') \frac{\delta \tau(\rv')}{\delta n(\rv)} d\rv',
\label{vxc}
\end{equation}
\begin{widetext}
\begin{eqnarray}
f_{xc}(\rv,\rv') &=&
\frac{\pa^2 \epsilon_{xc}}{\pa n^2}(\rv) \delta(\rv-\rv')  
-\left[ \nabla  \frac{\pa^2 \epsilon_{xc}}{\pa n \pa \nabla n}(\rv)\right]  \delta(\rv-\rv') 
- \nabla_i   \frac{\pa^2 \epsilon_{xc}}{\pa \nabla_i n \pa \nabla_j n}(\rv) \nabla_j \delta(\rv-\rv') \cr\cr
&+& \frac{\pa^2 \epsilon_{xc}}{\pa n \pa \tau} (\rv) \frac{\delta \tau(\rv)}{\delta n(\rv')}
+ \frac{\pa^2 \epsilon_{xc}}{\pa n \pa \tau} (\rv') \frac{\delta \tau(\rv')}{\delta n(\rv)}
- \nabla \frac{\pa^2 \epsilon_{xc}}{\pa \nabla n \pa \tau} (\rv) \frac{\delta \tau(\rv)}{\delta n(\rv')} 
-\nabla' \frac{\pa^2 \epsilon_{xc}}{\pa \nabla' n \pa \tau} (\rv') \frac{\delta \tau(\rv')}{\delta n(\rv)}
\cr\cr 
&+& \int \frac{\pa^2 \epsilon_{xc}}{\pa \tau^2}(\rv'') \frac{\delta \tau(\rv'')}{\delta n(\rv)} \frac{\delta \tau(\rv'')}{\delta n(\rv')} d\rv''
+\int \frac{\pa \epsilon_{xc}}{\pa \tau}(\rv'')\frac{\delta^2 \tau(\rv'')}{\delta n(\rv)\delta n(\rv')} d\rv''\!. 
\label{fxc}
\end{eqnarray}
\end{widetext}
The xc potential of Eq.~(\ref{vxc}) has been thoroughly addressed in Ref.~\onlinecite{Arbuznikov-03}
and our focus will be  the xc kernel of Eq.~(\ref{fxc}).
With the use of the standard perturbation theory, the functional derivatives of $\tau$ evaluate to   
\footnote{See EPAPS Document No  .}
\newcounter{ft}\setcounter{ft}{\value{footnote}}
\begin{eqnarray}
\frac{\delta \tau(\rv)}{\delta n(\rv')} \!  &=& \!
-v_s(\rv) \delta(\rv-\rv') \cr\cr
&+& \int H(\rv,\rv'') \chi_s^{-1}(\rv'',\rv') d\rv'' +\frac{1}4 \nabla^2 \delta(\rv-\rv'),
\label{dtdn} \ \ \  
\end{eqnarray}
\begin{eqnarray}
\frac{\delta^2 \tau(\rv)}{\delta n(\rv') \delta n(\rv'')}   &=& \! \!
-\delta(\rv \! - \! \rv')\chi_s^{-1}(\rv,\rv'')  \! - \! \delta(\rv \! - \! \rv'') \chi_s^{-1}(\rv,\rv') \cr\cr
&+& \! \! 2 \! \! \int \! \! \!F(\rv,\rv_1,\!\rv_2) \chi_s^{\!-1}(\rv_1,\!\rv') \chi_s^{\!-1}(\rv_2,\!\rv'') d\rv_1 d\rv_2\cr\cr
&+& \int H(\rv,\rv_1)  \chi_{s2}^{-1}(\rv_1,\rv',\rv'') d\rv_1,
\label{d2tdn2}
\end{eqnarray}
where
\begin{equation}
H(\rv,\rv_1) \! =
\! \!  \frac{1}{2} \! \sum\limits_{\alpha\ne \beta} \! \! \frac{(f_\alpha\! - \! f_\beta)(\epsilon_\alpha \! + \! \epsilon_\beta)}{\epsilon_\alpha \! - \! \epsilon_\beta} 
\psi_\alpha^*(\rv) \psi_\beta(\rv) \psi_\alpha(\rv_1) \psi_\beta^*(\rv_1), 
\label{H}
\end{equation}
\begin{eqnarray}
F(\rv,\rv_1,\rv_2) &=& \! \! \! \! \! \! \! \!
\sum\limits_{\alpha\ne\beta\ne\gamma\ne\alpha} 
\frac{f_\alpha\epsilon_\alpha-f_\beta\epsilon_\beta}
{(\epsilon_\alpha-\epsilon_\beta) (\epsilon_\alpha -\epsilon_\gamma)}  
\left[\psi^*_\alpha(\rv_2) \psi_\gamma(\rv_2) \right. \cr\cr
 &\times& \left. \psi^*_\gamma(\rv)
\psi_\beta(\rv) \psi_\alpha(\rv_1) \psi_\beta^*(\rv_1)  +(\rv\leftrightarrow \rv_1) \right] \cr\cr
&-&
\sum\limits_{\alpha\ne \beta}\frac{f_\alpha-f_\beta}
{(\epsilon_\alpha-\epsilon_\beta)^2} \epsilon_\beta
\left[  |\psi_\alpha(\rv_1)|^2 \psi^*_\alpha(\rv_2) \psi_\beta(\rv_2)  \right. \cr\cr 
&\times&
\psi_\alpha(\rv) \psi_\beta^*(\rv) \left. 
+ (\rv\leftrightarrow \rv_1) + (\rv_1 \leftrightarrow \rv_2) \right],
\label{2}
\end{eqnarray}
and  $\chi_{s2}^{-1}$ in Eq.~(\ref{d2tdn2}) is the inverse of the quadratic KS density-response function 
$\chi_{s2}(\rv,\rv',\rv'')= \frac{\delta^2 n(\rv)}{\delta v_s(\rv') \delta v_s(\rv'')}$.

Equations  (\ref{fxc})-(\ref{2})
together with the explicit KS response functions \footnotemark[\value{ft}]
constitute the complete solution to the MGGA-based xc kernel $f_{xc}$ (in the adiabatic approximation).

{\it Ultranonlocality --} 
In  reciprocal space, the xc kernel 
becomes a matrix in the reciprocal vectors $f_{xc,\Gv\Gv'}$(\qv).
Whether or not the MGGA for $f_{xc}$ provides an improvement over conventional approximations
depends on the presence or absence, in the optical limit $\qv\rightarrow \0v$, of a singularity of the type
$f_{xc,\0v\0v}(\qv) \simeq \alpha/q^2$~\cite{Reining-02}.
Obviously, LDA and GGA  [the first three terms in Eq.~(\ref{fxc})] do not have such a singularity.
With the neglect of the non-singular terms in the matrix form of Eq.~(\ref{fxc}) and without the  ``local-field effects", we simplify $f_{xc}$ to \footnotemark[\value{ft}]
\begin{equation}
f_{xc,\Gv\Gv'}(\qv)\approx  - \overline{\frac{\pa \epsilon_{xc}}{\pa \tau}} \ \chi_{s,\Gv\Gv'}^{-1}(\qv),
\label{approx2}
\end{equation}
where the overline denotes the average over the unit cell.  
The right-hand side of Eq.~(\ref{approx2}) {\em contains the singularity in question} 
because $\chi_s^{-1}$ does \cite{Pick-70,*Kim-02}.
Focusing on the ${\bf 0}{\bf0}$ component, we finally get
\begin{equation}
 \alpha=- \overline{\frac{\pa \epsilon_{xc}}{\pa \tau}} \lim_{\qv\rightarrow \0v} q^2 \chi_{s,\0v\0v}^{-1}(\qv).
\label{ouralpha}
\end{equation}
Considering that $\overline{\pa \epsilon_{xc}/\pa \tau}$ is almost the same for Si and Ge, neglecting for a moment the local-field effects,  and neglecting the unity compared to the static
dielectric function of a semiconductor, we see that Eq.~(\ref{ouralpha}) is in agreement with the empirical rule
of $\alpha$ being inverse-proportional to the dielectric function \cite{Botti-04}. 
We also note that the ultranonlocality we find
seems to be the first explicit demonstration of the fact that the kinetic energy-dependent functionals are not  in practice semi-local in the density \cite{Kummel-08}.

{\it Choice of functionals and calculation of optical properties -- }  
Having established on the fundamental level that the adiabatic meta-GGA-based TDDFT does account 
for the ultra-nonlocality in crystals, we now turn to numerical calculations.
First we note that only the group of functionals that provide $E_{xc}$ (e.g, VS98 \cite{Voorhis-98,*Voorhis-08} and
TPSS \cite{Tao-03})
rather than those providing $v_{xc}$ directly  (e.g., BJ06 \cite{Becke-06} and TB09  \cite{Tran-09}) can be used to build $f_{xc}$,
since for the functionals  of the latter group 
the corresponding $E_{xc}$ does not exist \cite{Tran-09}.
We have used two well established MGGA functionals VS98 \cite{Voorhis-98,*Voorhis-08} and TPSS \cite{Tao-03}  for the calculation of both the ground-state with $v_{xc}$ of Eq.~(\ref{vxc}) and $f_{xc}$ of Eq.~(\ref{approx2})
\footnote{For xc functionals,  calls to the Libxc subroutine library
(http://www.tddft.org/programs/octopus/wiki/index. \\ php/Libxc)
were used throughout.}. 
The resulting values of the key quantity 
$\overline{\pa \epsilon_{xc}/\pa \tau}$
entering Eq.~(\ref{approx2})  are listed in Table \ref{table}.
\begin{table}
\caption{\label{table}
The average over the unit cell of the derivative of the exchange, correlation, and xc energy density with respect to the kinetic energy density
found for Si and Ge with the VS98 \cite{Voorhis-98,*Voorhis-08}  and TPSS \cite{Tao-03} functionals. }
\begin{tabular}{llll|lllll}
\hline\hline
   & &  VS98 & & & TPSS  \\ \hline
\phantom{\Large A} & $\overline{\pa \epsilon_{x}/\pa \tau}$ & $\overline{\pa \epsilon_{c}/\pa \tau}$ & $\overline{\pa \epsilon_{xc}/\pa \tau}$ 
& $\overline{\pa \epsilon_{x}/\pa \tau}$ & $\overline{\pa \epsilon_{c}/\pa \tau}$ & $\overline{\pa \epsilon_{xc}/\pa \tau}$ 
\\ \hline
Si &  0.122 &-0.226 & -0.104 &  4.60$\times$10$^{-3}$ & -1.15$\times$10$^{-4}$ & 4.49$\times$10$^{-3}$\\ 
Ge &  0.135 & -0.241 & -0.106 & 2.94$\times$10$^{-3}$ & 8.73$\times$10$^{-5}$ & 3.03$\times$10$^{-3}$\\
\hline\hline
\end{tabular}
\label{tab}
\end{table}
At first glance surprisingly, the values found with the two different functionals differ drastically: The $\tau$-dependent  part of the TPSS functional was found negligible everywhere over the unit cell.  
In the Supplementary material \footnotemark[\value{ft}], 
we analyze the $\tau$-dependence of VS98 and TPSS functionals to the conclusion
that for the latter it is very weak. Accordingly, we argue that while well tuned
to yield accurate $E_{xc}$, TPSS performs unsatisfactorily with respect to its $\tau$-derivative. 
A clear reason for the weak $\tau$-dependence of TPSS  can then be easily identified: 
This functional is tuned to (i) the nearly free electron gas (NFEG) and (ii) the one and two electron systems \cite{Tao-03}. In both cases, due to the gradient expansion of the kinetic energy of NFEG
and to the von Weizs\"{a}cker's formula for the kinetic energy of one and two electron systems, respectively,
$\tau$ is (semi)-local in density, which leads to the local theory with respect to $f_{xc}$ and zero $\alpha$ (Cf. \cite{Nazarov-09}).
On the other hand, the VS98 functional is designed to work better in the strong  rather than  the weak
inhomogeneity case~\cite{Voorhis-98,*Voorhis-08}, which qualitatively explains its success in yielding realistic values of $\alpha$. Accordingly, we use the latter MGGA functional in our calculations below.

\begin{figure}[h] 
\includegraphics[width=1.0 \columnwidth]{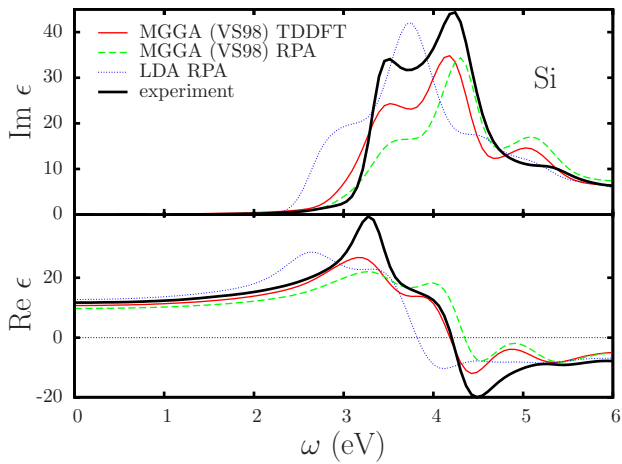} 
\caption{\label{fig_Si_ir} (color online)
Dielectric function of silicon.
Thin solid (red online) line is the  result obtained with MGGA band-structure and including  the many-body interactions
through  $f_{xc}$ of Eq.~(\ref{approx2}). 
Dashed (green online) line is the result obtained with MGGA band-structure but with $f_{xc}=0$ (RPA).
Dotted (blue online) line is obtained with LDA band-structure within RPA.
Thick solid line is the experiment from Ref. \onlinecite{Palik}.}
\end{figure}

\begin{figure}[h] 
\includegraphics[width=1.0\columnwidth]{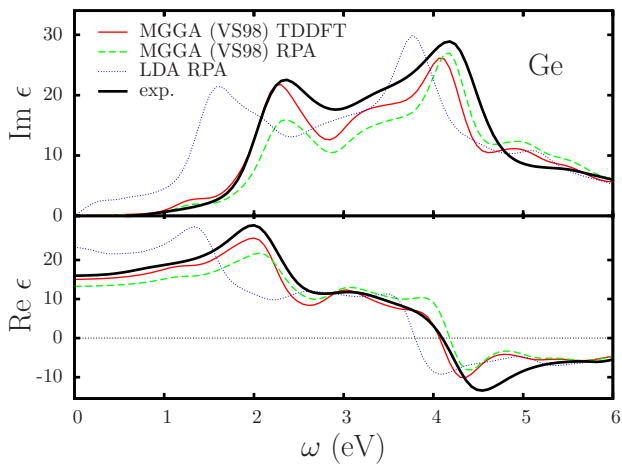} 
\caption{\label{fig_Ge_ir} (color online)
The same as Fig.~\ref{fig_Si_ir} but for germanium.}
\end{figure}

We calculated the KS band-structure and the microscopic density-response matrix of Si and Ge  with the full-potential linear augmented plane-wave (FP-LAPW) method and the VS98 MGGA xc functional
\footnote{The Elk FP-LAPW code (http://elk.sourceforge .net) with our implementation of MGGA xc functionals was used.}.
The supporting results for zincblende semiconductors are presented in \footnotemark[\value{ft}].
The real and imaginary parts of the macroscopic ($\qv=0$) dielectric function are presented in Figs.~\ref{fig_Si_ir} and \ref{fig_Ge_ir}.
It is evident that the inclusion of the
the non-local $f_{xc}$  of Eq.~(\ref{approx2})  via the MGGA greatly improves the agreement between the theory and experiment, in particular, making the excitonic peak considerably more pronounced. 

It is instructive to draw a parallel between our approach and that
of Ref.~\onlinecite{Reining-02}. Before the inclusion of $f_{xc}$, both methods produce  single-particle spectra
that underestimate  the intensity of the excitonic
peak. Then, with the inclusion of the many-body interactions through $f_{xc}$, the spectra are  red-shifted  and the excitonic feature grows.
The fundamental differences between the two approaches are that: (i) We remain all the time within the framework of TDDFT,
while Ref.~\onlinecite{Reining-02} uses a combination of TDDFT with the  GW approximation of many-body theory; 
(ii) While the quantity $\alpha$  in Ref.~\onlinecite{Reining-02} was a fitting parameter,
we have for it an explicit expression,  Eq.~(\ref{ouralpha}).
Moreover, for Si, Eq.~(\ref{ouralpha}) evaluates to $\alpha=-0.267$, which compares  reasonably well with the best fit value of  Ref.~\onlinecite{Reining-02} of $\alpha=-0.2$
\footnote{As VS98 contains empirical parameters, our numerical value of $\alpha$ cannot be viewed as {\em ab initio}.
We note, however, that the parameters' fit in Ref. \onlinecite{Voorhis-98} was completely unrelated to the excitonic effect.}.

In conclusion, we have developed the adiabatic TDDFT
formalism for the kinetic energy dependent (MGGA) exchange-correlation functionals.
In contrast to LDA and GGA approximations,
the resulting exchange-correlation kernel $f_{xc}$ is shown to exhibit the singularity
of the type $\alpha/q^2$, which is a necessary feature for a theory to describe the excitonic effect in crystals.
Our calculations performed for a number of the diamond-structure and zincblende semiconductors demonstrate the high promise of
the MGGA-based exchange-correlation functionals  as a new tool in the arsenal of TDDFT methods. 

\acknowledgements
VUN thanks E. E. Krasovskii for valuable discussions
and acknowledges partial support from National Science Council, Taiwan, Grant No. 100-2112-M-001-025-MY3.
GV acknowledges support from DOE Grant No. DEFG02-05ER46203.

%

\newpage
\newpage
\thispagestyle{empty}

\
\renewcommand{\theequation}{{A.\arabic{equation}}}
\renewcommand{\thefigure}{{A.\arabic{figure}}}
\onecolumngrid
\section{AUXILIARY MATERIAL\\
\ \\
to the paper by V. U. Nazarov and G. Vignale \\
\ \\
``Optics of semiconductors from the meta-GGA-based time-dependent density-functional theory''
}

\setcounter{page}{1}
\setcounter{equation}{0}
\setcounter{figure}{0}

\subsection{1. Derivation of Eqs.~(\ref{fxc})-(\ref{2})}
To evaluate the  functional derivatives of $\tau$ explicitly, we  note that the standard perturbation
theory  yields

\begin{eqnarray}
 &&\frac{\delta \epsilon_\alpha}{\delta v_s(\rv)}=|\psi_\alpha(\rv)|^2, 
\label{L1} \\
 &&\frac{\delta \psi_\alpha(\rv)}{\delta v_s(\rv')}=
\sum\limits_{\beta\ne\alpha} \frac{\psi_\alpha(\rv') \psi_\beta^*(\rv') \psi_\beta(\rv)}{\epsilon_\alpha -\epsilon_\beta},
\label{L2}
\end{eqnarray}
and by virtue of the functional chain rule 
 
\begin{eqnarray}
\frac{\delta \tau(\rv)}{\delta n(\rv')} \! = \! \int \frac{\delta \tau(\rv)}{\delta v_s(\rv'')} \frac{\delta v_s(\rv'')}{\delta n(\rv')} d\rv''
\! = \! \int \! \frac{\delta \tau(\rv)}{\delta v_s(\rv'')} \chi_s^{-1}(\rv'',\rv') d\rv'',
\end{eqnarray}
from Eq.~(\ref{t}), using  Eqs.~(\ref{L1}) and (\ref{L2}), we straightforwardly (though, with a lengthy algebra)
arrive at Eqs.~(\ref{dtdn}) and (\ref{d2tdn2}).
We also mention the properties
\begin{eqnarray}
&&\int H(\rv,\rv_1) d\rv = \int H(\rv,\rv_1) d\rv_1  =0,
\label{pH}\\
&&\int F(\rv,\rv_1,\rv_2) d\rv = \chi_s^{-1}(\rv_1,\rv_2),
\label{pF}
\end{eqnarray}
which directly follow from the definitions (\ref{H}) and (\ref{2}) with account of the orthogonality of the KS orbitals,
and which will be used below.

\subsection{2. Explicit forms of the density-response functions}

In equation (\ref{d2tdn2}), $\chi_{s2}^{-1}$ is the inverse of the quadratic KS density-response function.
A convenient relation holds

\begin{eqnarray}
\chi_{s2}^{-1}(\rv,\rv',\rv'') = - \int \chi_s^{-1}(\rv,\rv_3) \chi_{s2}(\rv_3,\rv_4,\rv_5) 
  \chi_s^{-1}(\rv_4,\rv')  \chi_s^{-1}(\rv_5,\rv'') d\rv_3 d\rv_4 d\rv_5, 
\label{chi2m}
\end{eqnarray}
which can be easily proven by functional differentiation of the identity $\chi_s \chi_s^{-1}=1$.
Explicitly, in terms of the KS orbitals, eigenenergies, and the occupation numbers
\begin{eqnarray}
 \chi_s(\rv,\rv') = \sum\limits_{\alpha\ne \beta}\frac{f_\alpha -f_\beta}{\epsilon_\alpha-\epsilon_\beta} 
\psi_\alpha^*(\rv) \psi_\beta(\rv) \psi_\alpha(\rv') \psi_\beta^*(\rv'),
\end{eqnarray}
\begin{eqnarray}
&&\chi_{s2}(\rv,\rv',\rv'') =
\sum\limits_{ \alpha\ne \beta}\frac{f_\alpha -f_\beta}{(\epsilon_\alpha-\epsilon_\beta)^2} \left[ |\psi_\beta(\rv'')|^2 -|\psi_\alpha(\rv'')|^2  \right]
\psi_\alpha^*(\rv) \psi_\beta(\rv) \psi_\alpha(\rv') \psi_\beta^*(\rv')+ (\rv \leftrightarrow \rv'') + (\rv' \leftrightarrow \rv'') \cr\cr
&&+ \! \! \! \! \! \! \! \sum\limits_{\alpha\ne\beta\ne\gamma\ne\alpha}  \! \! \! \frac{f_\alpha -f_\beta}{(\epsilon_\alpha-\epsilon_\beta)(\epsilon_\alpha -\epsilon_\gamma)} \psi_\alpha^*(\rv'') \psi_\gamma(\rv'') \psi_\gamma^*(\rv)
 \psi_\beta(\rv) \psi_\alpha(\rv') \psi_\beta^*(\rv') \! + \!  (\rv \leftrightarrow \rv'') \! + \! (\rv \leftrightarrow \rv')  \! + \!
(\rv\rightarrow \rv''\rightarrow \rv'\rightarrow \rv). \ \
\end{eqnarray}

\subsection{3. Derivation of Eq.~(\ref{approx2})}
Although Eq.~(\ref{approx2}) can be obtained directly as an approximation to Eq.~(\ref{fxc}),
it is more transparent to take a step back to Eq.~(\ref{vxc}).
Neglecting all the Fourier coefficients of the function$\frac{\pa \epsilon_{xc}}{\pa \tau}(\rv)$ but the zeroth
(which is supported by our calculations for Si and Ge)
we have from Eq.~(\ref{vxc})

\begin{equation}
v_{xc}(\rv) \approx \frac{\pa \epsilon_{xc}}{\pa n}(\rv)
 -   \nabla \frac{\pa \epsilon_{xc}}{\pa \nabla n} (\rv)
  +  \overline{\frac{\pa \epsilon_{xc}}{\pa \tau}} \int  \frac{\delta \tau(\rv')}{\delta n(\rv)} d\rv'.
\label{vxcapprox}
\end{equation}

Using Eq.~(\ref{dtdn}) and the property (\ref{pH}) of the function $H$ of Eq.~(\ref{H}), we can write
\begin{equation}
\int  \frac{\delta \tau(\rv')}{\delta n(\rv)} d\rv'=  
  - v_s(\rv).
\label{inttau}
\end{equation}
Another (instructive) way to prove Eq.~(\ref{inttau}) is to note that by the minimum principle
\begin{equation}
 \frac{\delta E}{\delta n(\rv)} = 0 = \left[\frac{\delta }{\delta n(\rv)} \int \tau(\rv') d\rv' \right] + v_{ext}(\rv) + v_{H}(\rv) + v_{xc}(\rv),
\label{min}
\end{equation}
where $E$ is the energy of the system. Equation (\ref{min}) immediately yields Eq.~(\ref{inttau}).

Substituting Eq.~(\ref{inttau}) into Eq.~(\ref{vxcapprox}) and taking the functional derivative of $v_{xc}(\rv)$ with respect to the density,
we arrive at

\begin{equation}
f_{xc}(\rv,\rv') \approx 
\frac{\delta}{\delta n(\rv')}
\left[
\frac{\pa \epsilon_{xc}}{\pa n}(\rv)
 -   \nabla \frac{\pa \epsilon_{xc}}{\pa \nabla n} (\rv)
\right]
 - v_s(\rv) \frac{\delta}{\delta n(\rv')} \overline{\frac{\pa \epsilon_{xc}}{\pa \tau}} 
-  \overline{\frac{\pa \epsilon_{xc}}{\pa \tau}} \frac{\delta v_s(\rv)}{\delta n(\rv')}.
\label{fxcapprox2}
\end{equation}

Finally, with the use of Eq.~(\ref{dtdn}), it is straightforward to show that all the terms in Eq.~(\ref{fxcapprox2}) but the last
are not singular as $1/q^2$ in the reciprocal space. By neglecting those terms and recalling the relation
\begin {equation}
 \chi_s^{-1}(\rv,\rv')= \frac{\delta v_s(\rv)}{\delta n(\rv')},
\end {equation}
we conclude the derivation of Eq.~(\ref{approx2}).
\vspace{-0.5 cm}
\subsection{4. Comparison between VS98 and TPSS}
\begin{figure}[h]
\includegraphics[height=0.45 \columnwidth]{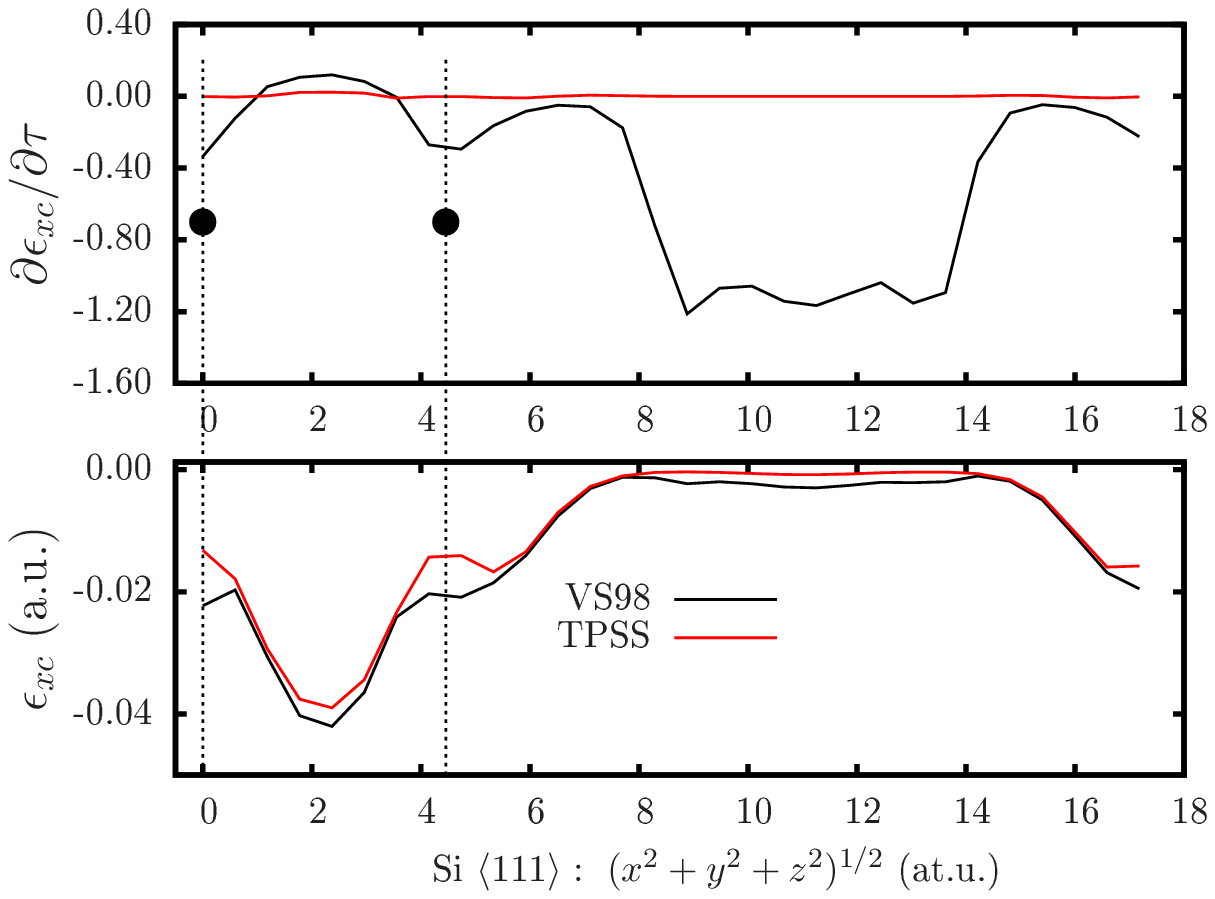}
\vspace{-0.25 cm}
\caption{The xc energy-density (lower panel) and its $\tau$ derivative (upper panel) along the $\langle 111 \rangle$ direction
of the cubic unit cell of Si crystal. In the case of TPSS functional (red line), $\pa \epsilon_{xc}/\pa \tau$ is negligibly small.
Circles show Si atoms positions.
\label{vtau}}
\vspace{-5 cm}
\end{figure}

\vspace{-0.25 cm}
Figure~\ref{vtau} demonstrates that while producing almost the same xc energy density in crystalline silicon (lower panel),
VS98 and TPSS functionals do it with very different 'weights' of the dependence on $n$ and $\nabla n$ on one hand,
and on $\tau$, on the other. In particular, TPSS functional exhibits almost no dependence on $\tau$ in the range of parameters
of this system (upper panel), hence, doing its job of approximating $\epsilon_{xc}$ by means of $n$ and $\nabla n$ only.
In other words, TPSS functional is almost a GGA one within this range of parameters.
On the contrary, VS98 functional depends on $\tau$ crucially, hence, being really nonlocal with respect to $n$.
\vspace{-0.25 cm}
\newpage
\subsection{5. Results for more semiconductors} 
Using Eq.~(\ref{approx2}), we have also performed calculations for zincblende semiconductors with the 
results shown in Fig.~\ref{fig_ir_TB09}.
To facilitate the calculations and to include $f_{xc}$ on top of the
as accurate as possible fundamental gaps, 
in the ground-state calculations we have used the highly efficient and accurate Tran and Blaha xc potential (TB09) \cite{Tran-09} while $f_{xc}$ has been included with VS98 as before. 
In order to demonstrate that the differences with the consistently used VS98 throughout (Figs.~\ref{fig_Si_ir} and \ref{fig_Ge_ir})  are minimal,
we also present results for Si and Ge obtained by this
slightly simplified method. We note that TB09 MGGA xc potential cannot be used for constructing
$f_{xc}$ since it gives the xc potential directly which does not correspond to any xc energy-density \cite{Tran-09}.

\newcommand{\s}{0.32}

\begin{figure}[h] 
\includegraphics[width=\s \columnwidth]{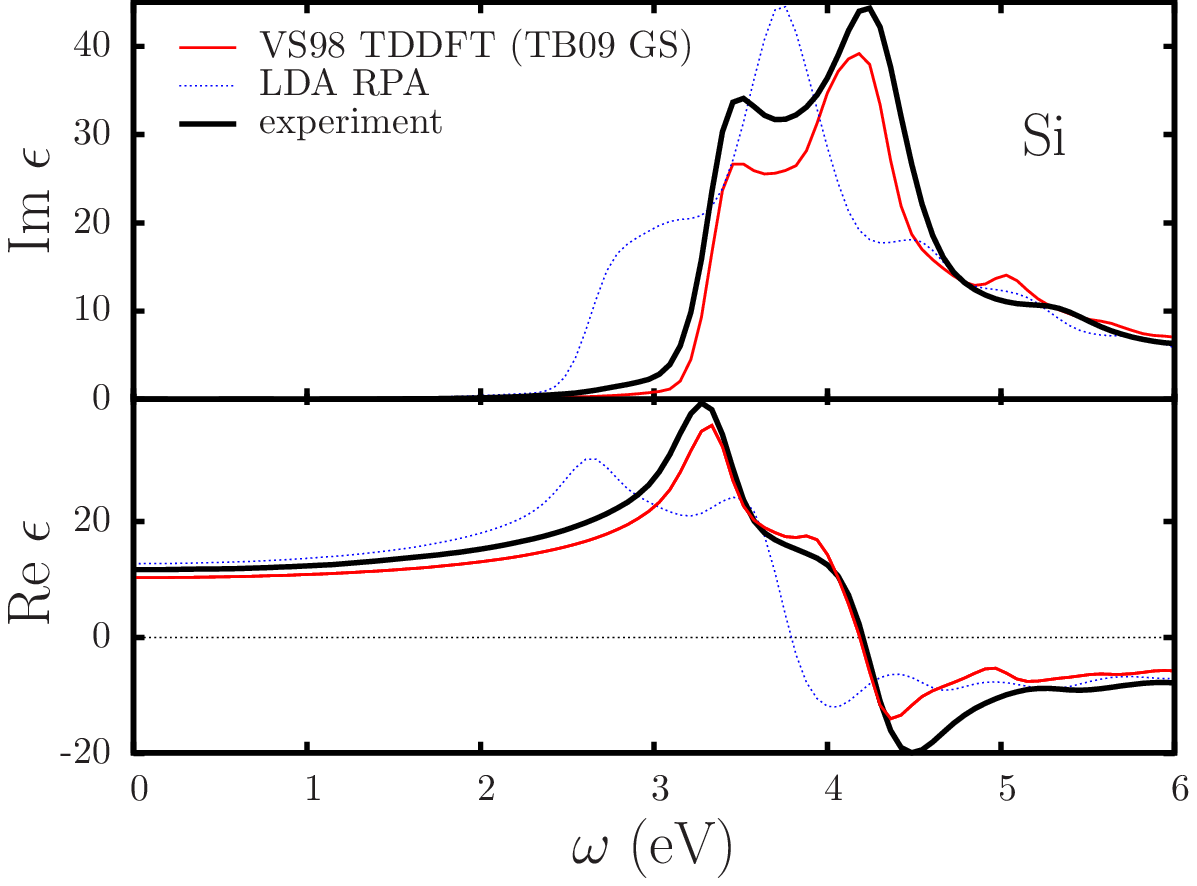} 
\includegraphics[width=\s \columnwidth]{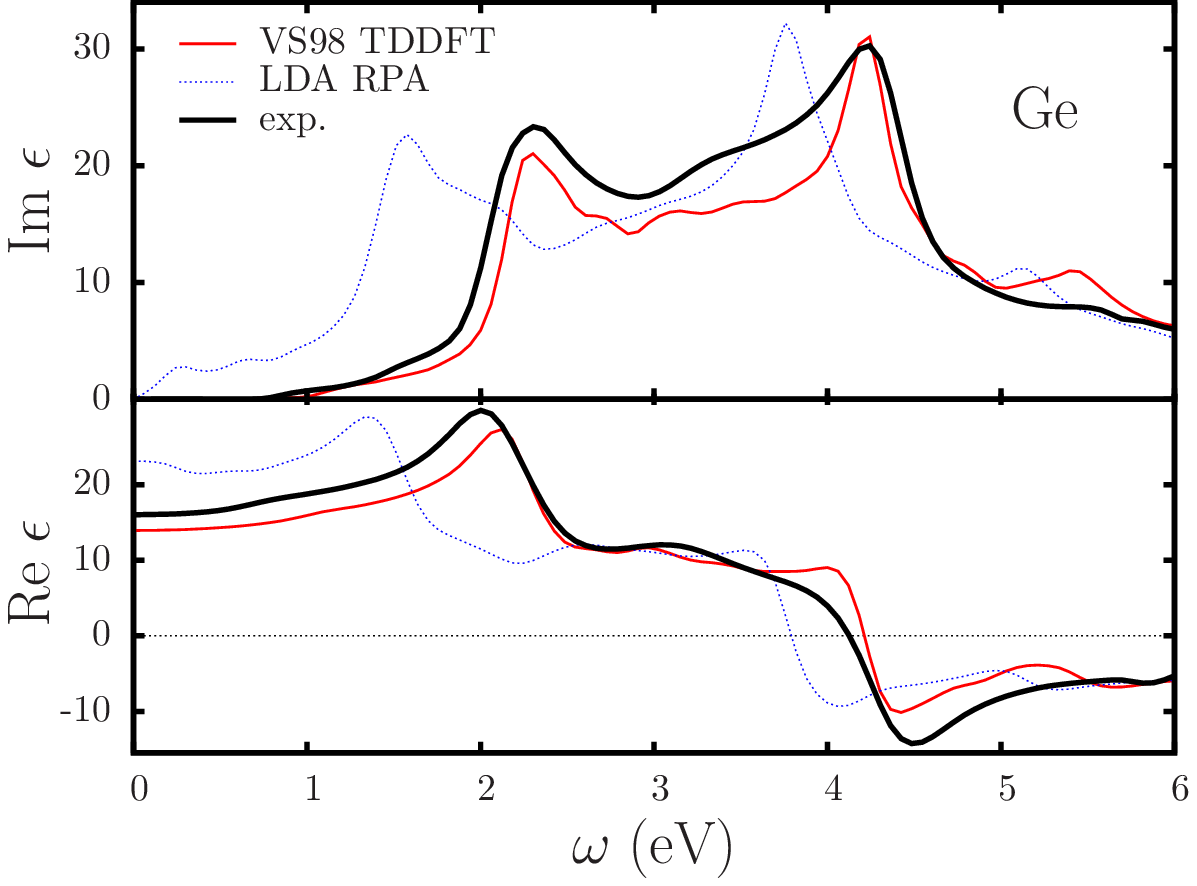} \\
\includegraphics[width=\s \columnwidth]{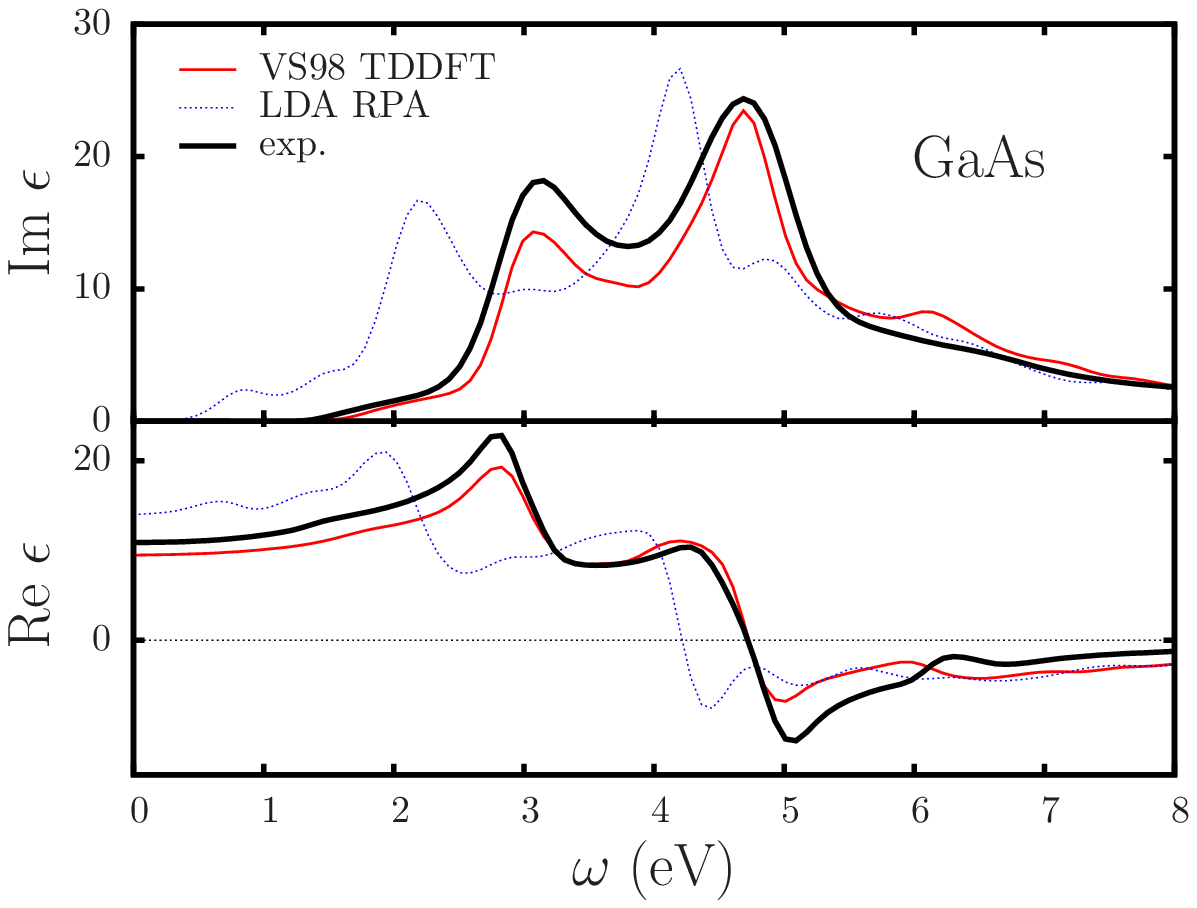} 
\includegraphics[width=\s \columnwidth]{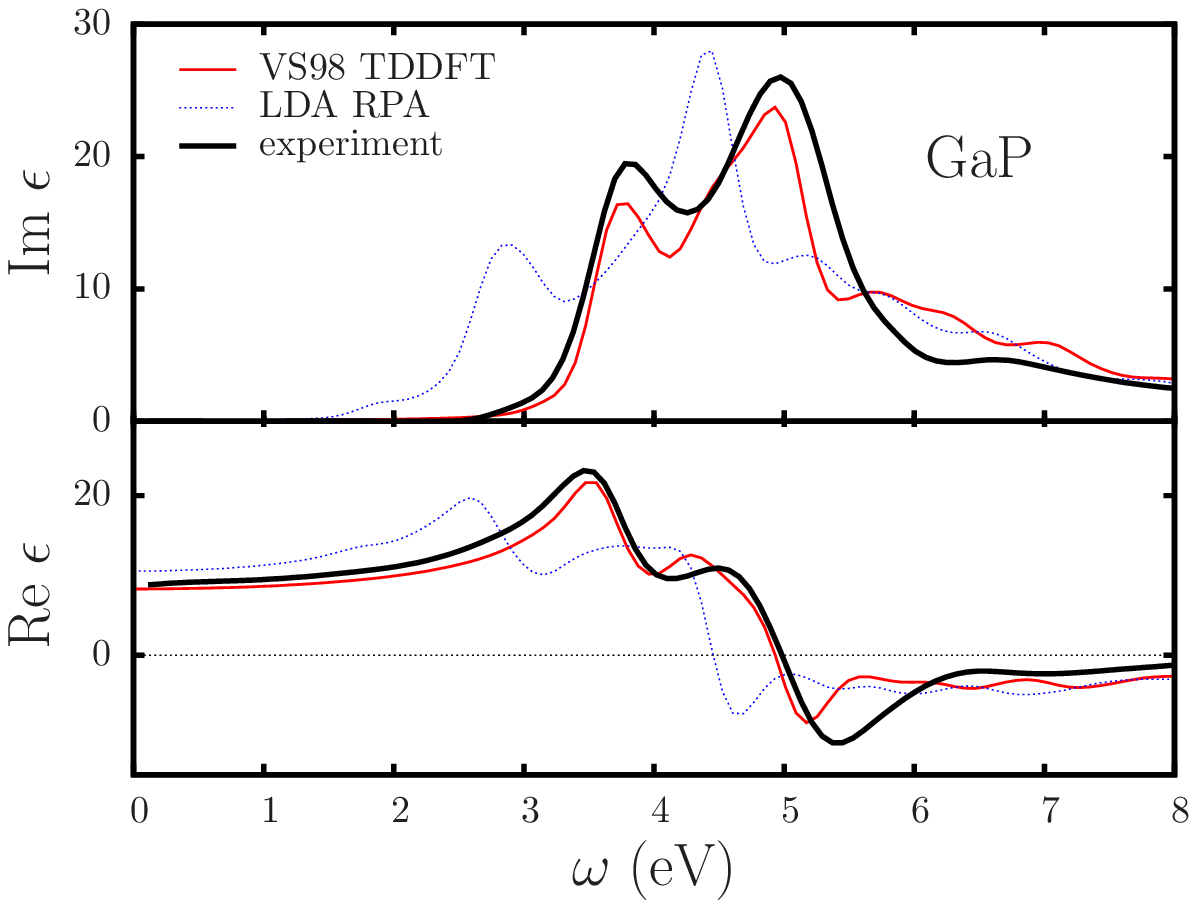} 
\includegraphics[width=\s \columnwidth]{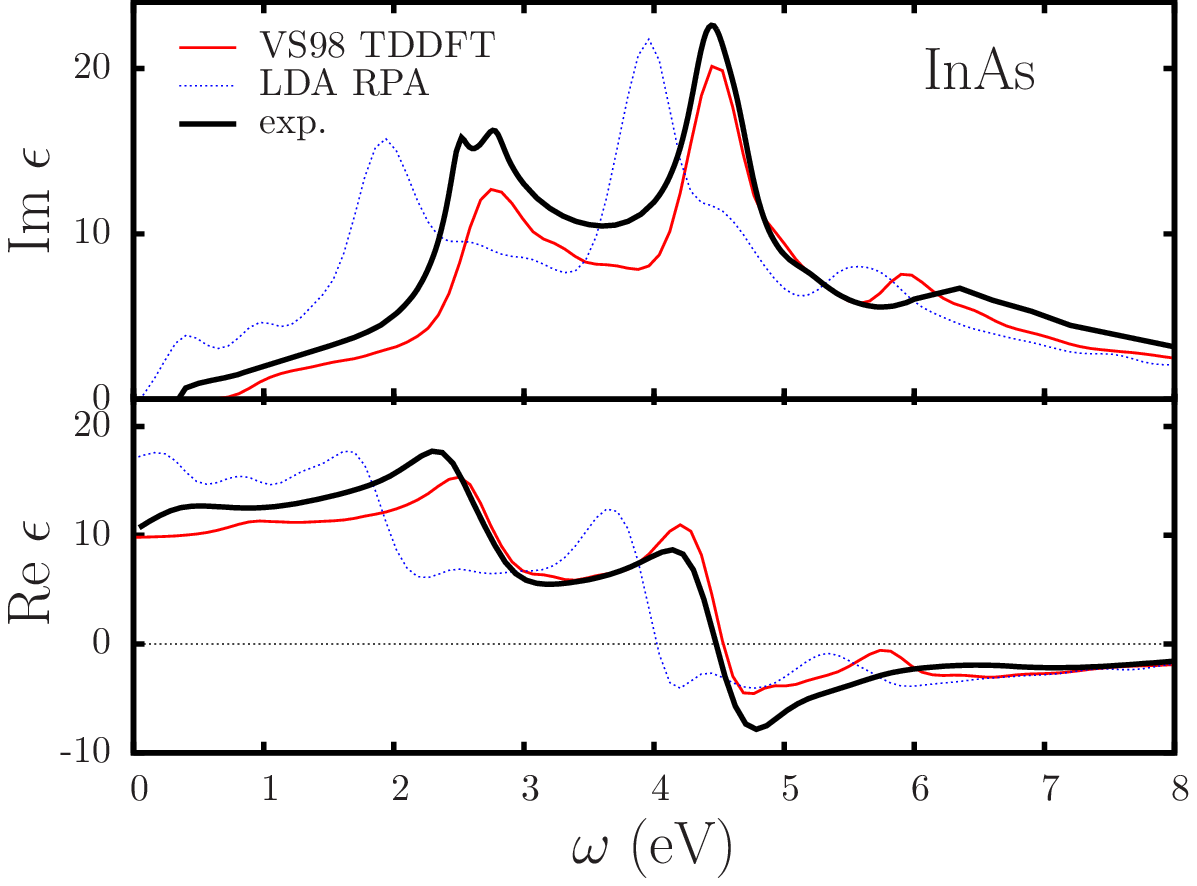} 
\caption{\label{fig_ir_TB09}
Dielectric function of Si, Ge, GaAs, GaP, and InAs semiconductors.
Red line is the  result obtained with the inclusion of  the many-body interactions
through  $f_{xc}$ of Eq.~(\ref{approx2}) using the VS98 MGGA xc functional while the ground-state band structure was obtained with TB09 xc potential. 
Blue line is obtained with LDA band-structure within RPA.
Thick solid line is the experiment from Ref. \onlinecite{Palik}.}
\end{figure}
\end{document}